\documentstyle[aps,manuscript]{revtex}

\newcommand{\be}{\begin{equation}}
\newcommand{\ee}{\end{equation}}
\def\la{\mathrel{\mathpalette\fun <}}
\def\ga{\mathrel{\mathpalette\fun >}}
\def\fun#1#2{\lower3.6pt\vbox{\baselineskip0pt\lineskip.9pt
\ialign{$\mathsurround=0pt#1\hfil##\hfil$\crcr#2\crcr\sim\crcr}}}
\begin{document}

\def\baselinestretch{1.5}
\normalsize
\vspace{2cm}
\title{Reply Comment on "Meson Masses in Nuclear Matter"}
\author{V. L. Eletsky$^{1,2}$ and B. L. Ioffe$^{1}$}
\address{$^1$Institute of Theoretical
and Experimental Physics,
 B.Cheremushkinskaya 25, Moscow 117259, Russia}
 \address{$^2$Institut f\"ur Theoretische Physik III, Universit\"at
Erlangen-N\"urnberg, D-91058 Erlangen, Germany}

\maketitle

\newpage
\def\baselinestretch{1.5}
\normalsize

As is well known, the operator product expansion (OPE) and QCD sum rule
method can be used in QCD for problems characterized by
short distances much smaller than the QCD scale, $l \ll \Lambda
\sim 0.5$ fm. Under no circumstances can these methods be exploited to treat long
distance problems, $l \ga 1$ fm. As was shown in our letter, meson (and baryon) 
mass shifts $\Delta m$ in nuclear matter are determined by
long distances. This follows from the basic relation of our letter

\be
\Delta m (E) = -2\pi\frac{\rho}{m}{\rm Re} f(E)
\label{dm}
\ee
in which the mass shift $\Delta m$ is expressed in terms of the real part of the meson-nucleon 
forward scattering amplitude, ${\rm Re} f$, and where $\rho$ is the nuclear density. 
We showed that for $\rho$-mesons $|{\rm Re} f| > 1$ fm. The same is true for $\pi$-mesons
at $E > 170$ MeV.  
If it were possible to determine $\Delta m$ in the short distance expansion, 
then ${\rm Re} f\ga 1$ fm could also be obtained in this way, which is obviously 
not true. It must be stressed, that Eq. (1) is valid
starting from very low energies. The only restriction comes from the
requirement that the particle wavelength be less than the internucleon
distance $d$. For $\rho$-mesons and normal nuclear density this restriction 
gives $E_{\rho}^{kin} > 10$ MeV. 
In our consideration of the $\rho$-meson mass shift we confined ourselves to high
energies, $E_{\rho}^{kin} \ga 1.2$ GeV,  only because we could not find
${\rm Re} f_{\rho N}$ at lower energies using the vector dominance model. 
(This restriction is absent in the pion case where we use experimental data on 
${\rm Re} f_{\pi N}$). Since the region $E_{\rho}^{kin} < 1$ GeV
is the resonance region in $\rho N$-scattering, we do not believe that, 
in this region and near the threshold, ${\rm Re} f_{\rho N}$ is much less than 1 fm. 
However, should it somehow happen to be the case (with short distance methods becoming
applicable), then we would get $|\Delta m_{\rho}|\la 10$ MeV, 
which still would not agree with the sum rule results of Hatsuda and Lee.

Let us remark that the two versions of OPE mentioned by Hatsuda and Lee, the 
light-cone (LC) expansion and the short distance (SD) expansion, are not in fact 
two different options. The second is a limiting case of the first. In the
space-time picture both correspond to expansion near the light cone, the 
second being done in the vicinity of its tip. 
Which version to choose is not in our hands - it is dictated by the physics
of the problem under consideration. In the case of a two-body problem 
(and the meson mass shift in nuclear matter is a two- or even many-body problem) 
the longitudinal distances along the light cone are of order $l \sim 1/mx$, 
where $m$ is the nucleon mass and $x = Q^2/2\nu$  is the Bjorken variable. 
In case of a $\rho$-meson at
rest $Q^2 \sim m^2_{\rho}$, $\nu \sim m m_{\rho}$, and $l$ is rather large, 
$l \sim 2/m_{\rho} \approx 0.5$ fm. Therefore, the SD expansion is not applicable here. 
It must also be mentioned, that in using fixed-${\bf q}$ dispersion relations, 
like the one in Eq. (3) of the Comment, a serious problem arises in the
phenomenological part of the sum rule. One must separate the contributions of
states in the meson channel which are of interest from the contributions of
baryonic resonances in $s$-  and, especialy, $u$-channels. The results of such 
separation are strongly model-dependent.

The statement made in the Comment that we overlooked in our letter the mean-field (MF) 
description of low-${\bf q}$ mesons is due to a misunderstanding which 
perhaps stems from the fact that we did not use the term "mean field theory" explicitly. 
(However, this term was used in Refs. [16,17] we quoted in our derivation). In fact,
our basic formula, Eq. (1), is a direct consequence of the MF
theory. The derivation of this formula in our letter used MF concepts
formulated in coordinate space and not in, the more commonly used, momentum space. 
It can also be derived from Eq. (1) of the Comment by considering 
the pole contribution to the l.h.s. of that equation. In our case, when the
meson momentum is much larger than the nucleon momenta in matter, the r.h.s. of 
Eq. (1) of the Comment factorizes and reduces to the nucleon
density appearing in our Eq. (1). Of course, at a low meson momentum
one should integrate over the nucleon momenta taking into account that ${\rm Re} f(E)$ 
is a function of the total energy in the c.m. frame. Positions and properties of 
resonances in meson-nucleon scattering are very important in such a
calculation, although in other aspects it is trivial.
At very low meson momenta, $|{\bf q}|^{-1}\gg d$, the averaging over the 
Fermi sea does not make sense anymore, since interactions with more than one 
nucleon and screening effects become important, and the MF approximation in 
its traditional form (linear in the Fermi distribution) breaks down. These effects
are not accounted for in calculations based on the SD expansion. 
\end{document}